\documentclass[12pt]{article}
\usepackage{amssymb,amsmath}
\usepackage{epsfig}
\usepackage{epstopdf}
\usepackage{latexsym}
\usepackage{graphicx}
\usepackage{booktabs}

\usepackage[margin=20pt,small]{caption}
\usepackage{subcaption}

\usepackage[toc]{appendix}

\usepackage{color}
\usepackage{datetime}
\usepackage[
      colorlinks=false,
      linkcolor=darkblue,  
      urlcolor=blue,    
      filecolor=blue,     
      citecolor=red,
linktocpage=true,
      pdfstartview=FitV,
      bookmarksopen=true,
	  hidelinks
      ]{hyperref}

\ifpdf
\fi

\newcommand*{\boxedcolor}{red}
\makeatletter
\renewcommand{\boxed}[1]{\textcolor{\boxedcolor}{%
  \fbox{\normalcolor\m@th$\displaystyle#1$}}}
\makeatother

\definecolor{cardinal}{rgb}{0.6,0,0}
\definecolor{darkgreen}{rgb}{0,0.5,0}
\definecolor{golden}{rgb}{0.92, 0.7, 0}
\definecolor{midnight}{rgb}{0, 0, 0.5}
\definecolor{darkblue}{rgb}{0.2, 0, 0.8}

\def\be{\begin{equation}}
\def\ee{\end{equation}}

\def\scrip{{\cal I}^+}
\def\1p{{(1p)}}

\def\p0{\phi_0}

\def\be{\begin{equation}}
\def\ee{\end{equation}}
\def\beq{\begin{eqnarray}}
\def\eeq{\end{eqnarray}}
 
\def\Hc{{\cal H}}

\def\sf{}

\def\jf{}
\def\j2{}

\def\p0{\phi_0}

\def\z0{\zeta_0}

\def\ave{\vec \alpha}

\def\beq#1{{\bf \small {(#1)}}}

\begin{document}  

\begin{titlepage}

\begin{center} 

{\Large \bf Observational Implications of Fuzzball Formation}

\bigskip
\bigskip
\bigskip

{\it Dedicated to the memory of Pierre Binetruy who lastingly advanced the exploration of fundamental physics with gravitational waves both through his brilliant science and by his monumental efforts at providing the tools to carry out the exploration.}

\bigskip
\bigskip
\bigskip
\bigskip

{\bf Thomas Hertog$^1$ and James  Hartle$^{2,3}$}\\

\bigskip
$\ ^1$ {\it Institute for Theoretical Physics, KU Leuven, Celestijnenlaan 200D, B-3001 Leuven, Belgium}\\
$\ ^2$ {\it Santa Fe Institute, Santa Fe, New Mexico, USA}\\
$\ ^3$ {\it Department of Physics, University of California, Santa Barbara,  93106, USA}\\

\bigskip
\bigskip

\texttt{thomas.hertog@kuleuven.be, hartle@physics.ucsb.edu}

\bigskip
\bigskip
\bigskip

{\bf Abstract}
\end{center}
\noindent We consider the quantum dynamics of gravitational collapse in a model in which the wave function spreads out over a large ensemble of geometries as envisioned in the fuzzball proposal. We show that the probabilities of coarse-grained observables are highly peaked around the classical black hole values. By contrast, probabilities for finer-grained observables probing the neighbourhood of collapsed objects are more broadly distributed and no notion of `averaging' applies to them. This implies that the formation of fuzzballs gives rise to distinct observational signatures that are more significant than has hitherto been thought and may be tested against observations in the near future. We also predict a novel kind of gravitational wave burst associated with the spreading of the wave function in gravitational collapse.

\end{titlepage}


\section{Introduction}
\label{intro}
Our observations of the progress and outcome of gravitational collapse are largely confined to the classical spacetime geometry far from the collapse itself. At late times and large distances this spacetime is a solution the classical Einstein equation characterized by the values of the mass ($M)$ angular momentum ($J$) and other multipoles ($Q$) that might be observable together with outgoing gravitational and electromagnetic radiation. A classical collapse that proceeds far enough will produce an event horizon, Hawking radiation, and the well known black hole information paradox.

Classical spacetime is not a given in quantum gravity. In a quantum mechanical treatment of collapse classical behavior is a matter of quantum probabilities supplied by the quantum state ($\Psi$) and dynamics $(H)$. A quantum system behaves classically when the probability is high for suitably coarse-grained histories of its motion to exhibit correlations in time governed by deterministic, classical, dynamical laws. The classical behavior of the flight of a tennis ball, the orbit of the Earth, and the spacetime geometry of a collapsing star are all examples. 

The classical approximation to the quantum dynamics may hold only in patches of the configuration space on which the state lives. Hence it does not necessarily imply a global classical spacetime. Moreover, in a given classical patch, states $\Psi$ generally predict relative probabilities for an ensemble of possible classical histories ---  not just one history \cite{Hartle2008,Hartle2007}. In between patches classical predictability breaks down, but quantum evolution connects classical histories in different patches \cite{Hartle:2015bna}. 

Classical predictability can break down on scales well below the Planck scale, and with no breakdown in the classical equations of motion. A widely applicable mechanism for the breakdown of classicality starts with a wave packet peaked about one classical history in an ensemble of possible classical histories. As long as the wave function remains peaked it predicts classical evolution. But if it spreads over other members of the ensemble then classicality has broken down. 

The familiar example of barrier penetration illustrates this \cite{Hartle:2015bna}. Consider a single particle moving in one-dimension. Suppose its wave function is peaked along a classical trajectory heading toward a barrier. When it hits the barrier the wave function will spread over two classical histories --- the reflected history and the transmitted history.  Classicality has broken down at the barrier but quantum evolution by the Schr\"odinger equation still holds. The classical history makes a quantum transition from one classical history into two. Examples involving gravity where quantum effects are important include the amplification of quantum fluctuations in the early universe and cosmological evolution across bounces \cite{Hartle:2015bna,Bramberger:2017cgf}.

Gravitational collapse is plausibly analogous. Consider the gravitational collapse of a spherical shell of ordinary matter in an asymptotically flat spacetime. Classically the shell evolves along a history satisfying the Einstein equation. As the collapse proceeds an event horizon forms, and eventually a singularity inside the event horizon. Quantum mechanically this same collapse could be described by starting with a wave function satisfying the Wheeler-DeWitt equation that initially is peaked around the geometry, fields and their conjugate momenta of a classical solution of the Einstein equation. We would expect the wave packet to follow the classical solution at least for a while. We also expect the classical approximation to break down when the shell is sufficiently close to the classical singularity. But suppose that the wave function spreads out sufficiently early. Then classical evolution has broken down further from the singularity than suggested by naive Planck scale arguments. Instead, the breakdown is governed by the rate of spreading  -- often through tunneling. A breakdown sufficiently far out might not leave enough classical spacetime to define an event horizon globally as the boundary of the past of future null infinity. Then there would be no black hole information paradox \cite{Mathur:2009zs,Hartle:2015bna}.  

The fuzzball (FB) scenario\footnote{See e.g. \cite{Mathur:2005zp,Skenderis:2008qn,Balasubramanian:2008da,Bena:2013dka} for reviews of the fuzzball program.} is an example of a model of gravitational collapse that envisions an early spreading of the wave function of this kind \cite{Mathur:2008kg,Kraus:2015zda,Bena:2015dpt}. The fuzzball program has been developed in detail only for certain classes of BPS black holes, and mostly in five dimensions. However it has been argued that its key qualitative features may hold more generally, including in four dimensions. In this paper we assume this and explore the observational implications of a fuzzball-based model of gravitational collapse. We find that upcoming observations of the near horizon region have the potential to explore and constrain the FB scenario and related models of gravitational collapse exhibiting significant deviations from  classical black hole evolution in the near-horizon region.

The FB program rests on the observation that five-dimensional supergravity theories admit a large class of solutions that are smooth, geodesically complete, horizonless, asymptotically flat spacetimes with a complicated and extended geometric and topological structure in an interior region, but with the same asymptotic charges as (BPS) black holes. These fuzzball solutions or microstate geometries are conjectured to be semi-classical descriptions of black hole microstates\footnote{Five-dimensional supergravity circumvents no-go theorems in four dimensions that exclude the existence of such regular solitonic solutions because it has Chern-Simons interactions and the spatial sections may have non-trivial second homology \cite{Gibbons:2013tqa}. At first sight this suggests the FB model of gravitational collapse cannot hold in four dimensions. However there are richer classes of microstate solutions that are singular when considered as four-dimensional spacetimes but yield smooth solutions when lifted to higher dimensions \cite{Saxena:2005uk,Balasubramanian:2006gi,Denef:2007yt}.}. The FB scenario of gravitational collapse asserts that at approximately the Schwarzschild radius the wave function spreads over an ensemble of microstate geometries. The spreading can occur through tunneling to and between different microstate solutions. The end state of gravitational collapse in the FB program is thus given by a wave function that describes a superposition of geometries with the same mass and angular momentum as the pre-collapse configuration but differing in further multipole moments of geometry and fields. 

We use sum-over-histories quantum mechanics \cite{Hartle:1992as,Har93a} to construct a toy model of the quantum dynamics of gravitational collapse in a scenario like what is envisioned in the fuzzball proposal\footnote{See e.g. \cite{Giddings:2016btb,Giddings:2016tla,Cardoso:2016oxy, Abedi:2016hgu} for recent studies of possible quantum gravity signatures in alternative models of gravitational collapse.}. In this framework a given collapse history corresponds to one particular branching of the wave function and not some kind of average over an ensemble of fuzzballs. 
It involves a series of tunneling events which generally break the symmetry and are therefore expected to lead to bursts of gravitational and electromagnetic radiation. Individual fuzzballs being regular and asymmetric they necessarily differ significantly from black holes, and possibly from each other, on the scale of the would-be horizon. This implies that the quantum mechanical probabilities of sufficiently fine-grained observables probing the spacetime geometry in the neighbourhood of fuzzballs are expected to be relatively broadly distributed, thus providing observational discriminants between fuzzballs and black holes. Observations holding great promise to probe some of these distributions include not only those of gravitational wave patterns resulting from mergers \cite{TheLIGOScientific:2016src,Audley:2017drz} but also the upcoming images of the immediate environment of SgrA* and M87 with the Event Horizon Telescope \cite{Doeleman:2009te}. By contrast the probabilities of coarser-grained observables sensitive to the spacetime geometry at large distances only are highly peaked around their black hole values. Coarse-grained observations effectively see an average over fuzzball geometries and we argue it is at this level that classical black holes emerge in the theory.

\section{Fuzzball Formation}
\label{fuzzballs}

In a quantum theory of asymptotically flat spacetimes  states are described by wave functions $\Psi$ on the superspace of geometries and field configurations on a family of spacelike surfaces $\Sigma_t$ labeled by their asymptotic time $t$. This wave function obeys the Hamiltonian constraint of the theory which can be put in the form
\be
i\hbar \frac{\partial\Psi}{\partial t} = {\cal H} \Psi
\label{wdw}
\ee
where $\cal H$ is the Wheeler-DeWitt operator. 

Consider the gravitational collapse of a spherical shell of ordinary matter described in the Introduction.  Assume, as described there, that the collapse is described by starting with a wave function that initially is peaked around the geometry, fields and their conjugate momenta of a classical solution of the Einstein equation. And assume that this wave packet would follow the classical solution at least for a while.  

The premise of the FB program is that the wave function spreads through tunneling over an ensemble of horizonless FB solutions well before a singularity forms. The ensemble of FB solutions produced in a given gravitational collapse will have the same asymptotic charges as the black hole that would form were the collapse to proceed classically. When viewed from the outside, the fuzzballs differ from the black hole and from each other in the values of the higher multipole moments of geometry and fields. Evidently they differ drastically from the black hole in the interior region, being regular and exhibiting a complicated, extended geometric and topological structure involving one or more extra dimensions, etc.

Classical evolution as represented by the peaked wave packet breaks down in the fuzzball model of gravitational collapse earlier than what one might have anticipated on the basis of classical, four-dimensional general relativity. At intermediate times the wave function of the collapsing body describes, at an appropriate level of coarse-graining, an ensemble of alternative {\it quasi}-classical histories with high quantum probabilities for classical correlations over stretches of time interrupted by quantum events such as tunneling\footnote{This is not unlike the situation in false vacuum eternal inflation in cosmology where (at an appropriate level of coarse-graining) the wave function of the universe describes an ensemble of quasi-classical histories with bubbles nucleating at different times and different places. See e.g. \cite{Hartle:2016tpo} for a recent discussion of this.}. 

In sum-over-histories quantum mechanics \cite{Hartle:1992as,Har93a} the theory $(H,\Psi)$ predicts probabilities for which of a set of alternative decoherent histories happens in the universe. Such sets are called realms. Beyond the coarse-graining needed for decoherence there are sets of histories defined at various levels of further coarse graining, depending on which features or observables are followed and which are ignored. We are interested here in the particular quasi-classical realm that follows the quantum transitions that give rise to the spreading of the wave function but ignores possible other quantum effects.

In a non-relativistic approximation this set of alternative histories can be described by introducing at any one moment of time $t$ an exhaustive set of exclusive Heisenberg picture projection operators $\{P_{\alpha}(t)\}$, acting in $\Hc$, with $\alpha$ labeling the different tunneling channels. The operators satisfy 
\be
\label{projections1}
P_{\alpha}(t)P_{\alpha'}(t) = \delta_{\alpha \alpha'} P_{\alpha}(t), \quad \sum_{\alpha} P_{\alpha}(t) = I . 
\ee
Projection operators representing the same quantity at different times are connected by unitary evolution by the Hamiltonian $H$
\be
P_\alpha(t') = e^{iH(t'-t)}P_\alpha(t) e^{-iH(t'-t)} . 
\label{un-evol}
\ee
A set of alternative coarse-grained histories is specified by a sequence of such sets at a series of times $t_1,t_2, \cdots t_n$. An individual history corresponds to a particular sequence of events $\vec \alpha \equiv (\alpha_1,\alpha_2, \cdots, \alpha_n)$ and is represented by the corresponding chain of projections viz:
\be
C_{\vec \alpha} \equiv P^n_{\alpha_n}(t_n) \dots P^1_{\alpha_1}(t_1) .
\label{class-closed}
\ee
Such sets of histories are coarse-grained, not only because the $P$'s are coarse grained,  but because they occur at some times and not at all times.  Branch state vectors corresponding to the individual quasi-classical histories that keep track of the sequence of tunneling transitions can be defined by
\be
\label{branch}
|\Psi_{\ave} \rangle \equiv  C_{\ave} |\Psi\rangle . 
\ee
The set of histories decoheres when the quantum interference between branches is negligible
\be
\label{decoherence}
\langle\Psi_{\ave} |\Psi_{\vec \beta}\rangle \propto \delta_{\ave{\vec \beta}}.
\ee
As a consequence of decoherence consistent probabilities $p(\ave)$ can be assigned to the individual histories that are 
\be
\label{probs}
p(\ave) = |||\Psi_{\ave}\rangle||^2 = ||C_{\ave} |\Psi\rangle||^2.
\ee
A decoherent set of alternative coarse-grained histories is called a `realm' for short.  

Tunneling transitions between different FB solutions can be described at this level of coarse graining and in this  approximation by considering the two time history in which the system is in FB state $\alpha$ at a time $t$ and in different FB state $\alpha'$ a time $\Delta t$ later. From \eqref{probs} the probability for this history, including the quantum transition, is
\be
\label{transprob}
p(\alpha \rightarrow \alpha') = ||P_{\alpha'}(t+\Delta t)P_{\alpha}(t )|\Psi\rangle||^2.
\ee
When the probability for this transition is divided by $\Delta t$ this gives the semiclassical approximation to the nucleation rate between these FB states per unit time. 

\begin{figure}[ht]
\centering
\includegraphics[height=3.7in]{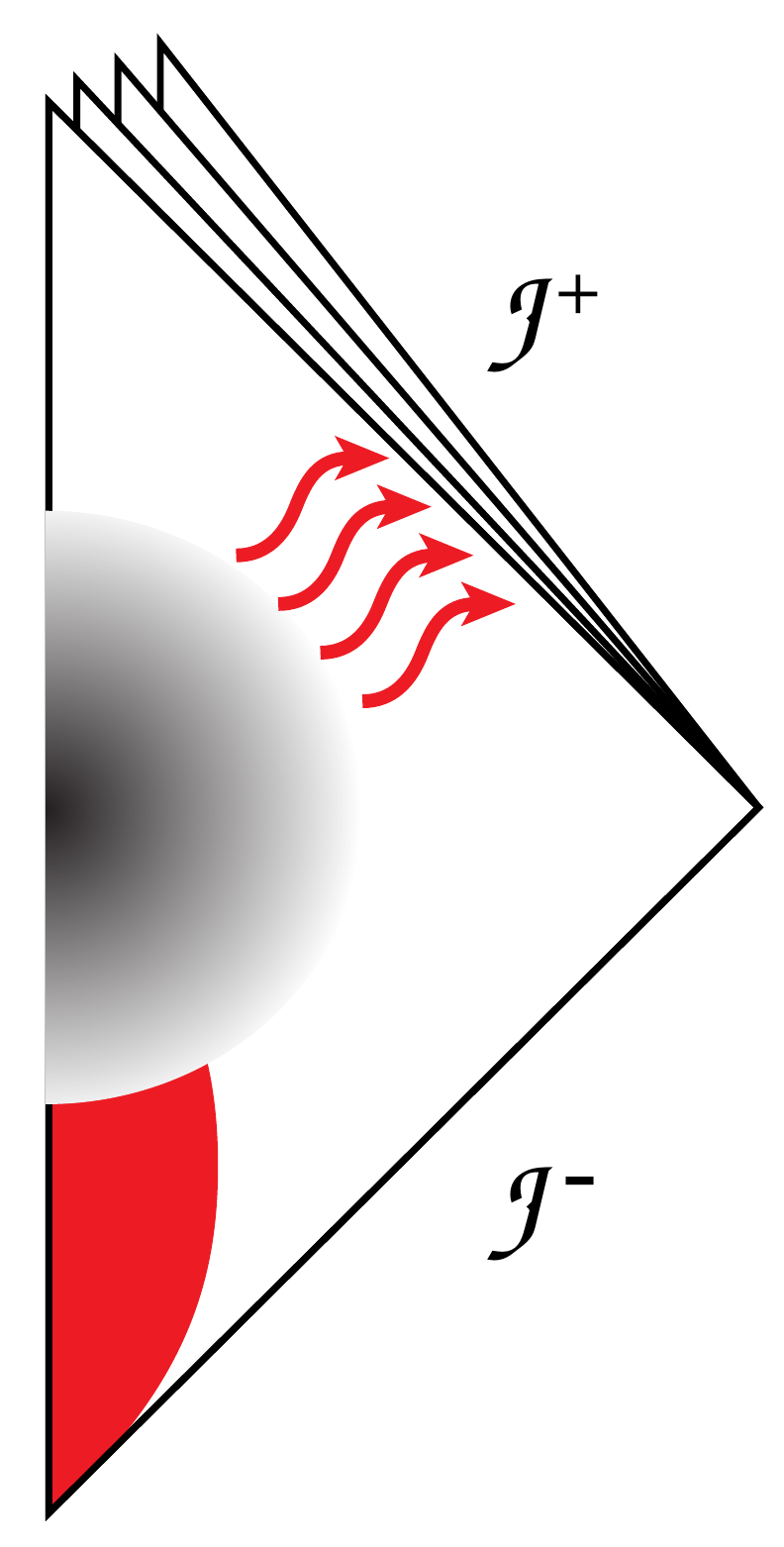} 
\caption{A schematic representation of the quantum dynamics of gravitational collapse in a model in which the wave function spreads over a large ensemble of classical endstates as envisioned in the fuzzball proposal. Classical evolution breaks down in an extended (shaded) region away from the classical singularity. Quantum evolution mediated by the Wheeler-DeWitt equation provides probabilities for quantum transitions of the initial classical spacetime to {\jf a set of} final classical spacetimes near $\scrip$, which in the fuzzball proposal differ for example in the multipole moments of geometry and fields. There is no notion of an absolute event horizon as the boundary of the past of $\cal I^+$ because there is not enough classical spacetime to define it. There is then no causal obstacle to information emerging, but the information is spread out over a range of final spacetimes.}
\label{bhclass-sptime}
\end{figure}

The tunneling rate between any two classical solutions is exponentially small. However in the FB program it is envisioned that this suppression is compensated for by the great number of different tunneling channels, rendering the total probability to tunnel of order one when the radius $R$ of a collapsing shell approaches $\sim 2M$ \cite{Mathur:2008kg,Kraus:2015zda}. This leads to the spreading of the wave function over an ensemble of FB states before a singularity forms. A schematic representation of this evolution is given in Figure \ref{bhclass-sptime}. The analogy with tunneling through barriers in quantum mechanics suggests that the tunneling rate decreases as the wave function further spreads, and that the collapsing system eventually settles down. 
The state at late times would then be given by a wave function that describes a stationary superposition over a sector of classical FB states.

\section{A Barrier Toy Model}
\label{barrier}

Transitions between classical histories can leave observable signatures. We illustrate this first with a simple generalization of the {\sf one-dimensional} barrier penetration example described in the Introduction to three dimensions $(x,y,z)$. Suppose that our particle has a spin $\vec s$ and a magnetic moment $\mu{\vec s}$. Suppose that the barrier is like a hill in $x$ but translationally invariant in the $y$ and $z$ directions. Suppose further that there is a magnetic field $\vec B$ in the $y$-direction in a region near the barrier on the far side. Assume initially that the particle is in a wave packet on one side of the barrier that is peaked about the position and momentum of a classical trajectory moving along the $x-$axis toward the barrier. Assume that initially the magnetic moment is up --- along the $z$-direction. 

After passing through the barrier the wave function will be in a superposition of two classical histories --- the reflected history and the transmitted history. The transmitted wave packet will be moving in the $+x$ direction with its magnetic moment changed by the magnetic field it has passed through and electromagnetic radiation emitted as a consequence. The reflected wave packet will be moving in the $-x$ direction with the magnetic moment unchanged and no emitted electromagnetic radiation.

We can make simple dimensional estimates of the amount of electromagnetic radiation emitted. We start with the familiar formula for the luminosity $L_{EM}$ from a time changing magnetic dipole
\be
\label{magdipole}
L_{EM} = \frac{c}{3} |\ddot{\vec\mu}|^2.
\ee
Suppose that $T$ is the time that the particle spends in the magnetic field under the barrier.  Then rough estimates for $L_{EM}$ and  for the total energy emitted $E_{EM}$ are
\be
\label{EMest}
L_{EM} \sim \frac{c {\vec\mu}^2}{3T^2}, \quad\quad E_{EM} \sim \frac{c{\vec\mu}^2}{3T}.
\ee

Suppose an observer examines this system. At a fine-grained level the observer could determine the particle's position, the value of the magnetic moment, and the emitted electromagnetic radiation. The observer will find values that characterize either the reflected or transmitted classical history not some average of them. Subsequent observations will be consistent with just one of the histories.
In the language of Copenhagen quantum mechanics we could say that the state has collapsed on one of the two possible classical histories. 

Suppose the observer was only able to measure the emitted radiation --- a much coarser grained description of the system. Given the state the observer would know that the particle was on the transmitted branch and predict what the outcome of finer grained measurements of position and magnetic moment would yield.

\section{Observational Implications of Fuzzballs}
\label{observations}

Descriptions of the outcomes of gravitational collapse are available at various levels of coarse graining. To discuss possible connections with observations we begin with a coarse enough graining that defines an ensemble of classical FB histories at late times. Beyond this the degree of coarse-graining is connected with  the observable being considered\footnote{This is a fine-graining compared to the black hole history but there is little evidence that the ensemble of regular FB solutions in supergravity is the maximally refined classical ensemble of histories describing the endstate of gravitational collapse. For instance classical string theory corrections may be a further refinement which has been argued to have important implications in generalizations of the FB proposal to realstic black holes. This also means that failure to account for the black hole entropy by counting fuzzball solutions only need not invalidate the fuzzball proposal. Rather what matters is whether the wave function at this level of coarse-graining has most of its support over FB states.}.

The conserved mass $M$ and angular momentum $\vec J$ of the initial configuration will of course be preserved in the tunneling transitions involved in the formation process of fuzzballs. But the quadrupole and higher moments of the asymptotic metric and fields will change in general, and one expects tunneling to go together with the emission of gravitational and electromagnetic radiation. Furthermore the pattern of radiation will depend on the tunneling channel. We are therefore led to consider two classes of observables ${\cal O}$ corresponding to two qualitatively different levels of coarse graining:

\subsection{Coarse-grained Observations}
\label{CG}
First we assume the system has settled down and we consider observables that probe only the conserved charges such as mass $M$ and angular momentum $\vec J$, defined at large distances from the collapsed object and specifying the asymptotic structure of the solutions only. The general expression for the probabilities of such observables ${\cal O}$ in the ensemble of fuzzball histories is given by
\be \label{obs}
p({\cal O}) = \sum_{\ave} p({\cal O}|\ave) p(\ave)
\ee
where $\ave$ labels the different fuzzball histories over which the initial wave function spreads. Observables with probability distributions that are sharply peaked around specific values are predicted by the theory with high confidence to take these values. 

The expression \eqref{obs} simplifies for coarse-grained observables ${\cal O}_{CG}^{\ }$ with probability distributions that are independent of the specific fuzzball history. Evidently this is the case for $M$ and $J$ which take the same values in all histories in the ensemble. For observables of this kind \eqref{obs} reduces to
\be \label{obscg}
p({\cal O}_{CG}^{\ }) = p({\cal O}_{CG}^{\ }) \sum_{\ave} p(\ave) = \delta ({\cal O}_{CG}^{\ } - {\cal O}_{cl}^{\ }).
\ee
Therefore to compute the predictions for coarse-grained observables one can bundle the individual fuzzball histories together into a much coarser-grained history that does not differentiate between the different possible sequences of tunneling events in the formation process. In this way a notion of `averaging' over the individual fuzzball histories emerges, with a measure provided by the theory. 
\subsection{Fine-grained Observations}
\label{FG}

Next we consider finer-grained observables that probe the spacetime and fields in the neighborhood of the collapsed object. Examples not only include observations of inspiraling stars or EMRI's and light rays moving around the gravitational collapsed object but also the detailed gravitational wave patterns resulting from mergers. Fine-grained observations of this kind enable one to determine not only the mass $M$ and angular momentum $J$ of the spacetime, but conceivably also some of the higher multipole moments $Q$. Were the collapse governed by general relativity in four dimensions the result would be a Kerr black hole in which the higher multipoles $Q$ are determined by $M$ and $J$. The conserved mass and angular momentum are determined by the initial condition but the relation $Q(M,J)$ is determined by the theory. 

Fuzzball microstate geometries differ significantly from black holes in the region where curvatures are $\sim M^{-2}$. For instance they neither have a horizon nor do they have the same symmetries as black holes. It is thus plausible that relatively fine-grained observations determining some number of multipoles and the emitted gravitational radiation in mergers should discriminate between different microstates. Specifically, the probability distributions \eqref{obs} for fine-grained observables ${\cal O}_{FG}^{\ }$ are given by
\be \label{obsfg}
p({\cal O}_{FG}^{\ }) = \sum_{\ave} p({\cal O}_{FG}^{\ }|\ave) p(\ave) 
\ee
where the distributions $p({\cal O}_{FG}^{\ }|\ave)$ peak at different values in different branches $\ave$, yielding a broader distributions $p({\cal O}_{FG}^{\ })$.

Hence detailed fine-grained observations of the collapsed object do not see an average over the fuzzball states. Instead an observer outside watching a given collapsed system will see fine-grained observations of outgoing radiation probing the neighbourhood of collapsed objects consistent with one member of the classical ensemble of histories. 
Observations of $Q$ that did not satisfy the constraint predicted by standard general relativity but were consistent with one of the other microstates would support the fuzzball scenario of gravitational collapse.
Future observations of gravitational waves e.g. of the ringdown phase after mergers will test the black hole no-hair theorems and therefore provide particularly promising examples of fine-grained observables ${\cal O}_{FG}^{\ }$ enabling precision tests of this kind \cite{TheLIGOScientific:2016src,Audley:2017drz}.

More importantly when it comes to testing the FB program are detailed observations of many collapse histories in many different locations in the universe. This would enable one to gather data enabling one to verify the statistical properties of the distributions \eqref {obsfg} predicted by the theory, such as its variance, non-Gaussianity in certain directions etc, thereby yielding a precision test of the fuzzball picture -- and hence of quantum gravity more generally.

\subsection{Gravitational Wave Bursts}
\label{gw}

Finally we move on to some of the observational implications of the fuzzball formation process itself. Fuzzball formation proceeds by quantum transitions between different classical geometries. In a transition between geometries that have different asymptotic multipole moments gravitational radiation can be expected. Simple dimensional estimates show that this could be observationally significant. 

For simplicity consider a collapse with a mass $M$ and zero total angular momentum. Assume that in making the quantum transition between one classical history and another the quadrupole moment changes by an amount $\Delta Q$ in a time $T$. The luminosity emitted in gravitational waves  $L_{GW}$ can be estimated by the quadrupole formula from linearized gravity 
\be
\label{quadform}
L_{GW} =\frac{1}{5}\frac{G}{c^5} (\dddot Q_{ij} \dddot Q^{ij}) .
\ee
Here $Q_{ij}$ is the (trace free) quadrupole moment tensor and a dot denotes a time derivative.  The total energy emitted in gravitational waves is then roughly,
\be
\label{totGW} 
E_{GW} \sim \frac{G}{c^5}  \left(\frac{\Delta Q}{T^3}\right)^2 T .
\ee
Write $\Delta Q =\eta_Q M R^2$, where $R$  is the gravitational radius  $GM/c^2$ and  $\eta_Q$ is a dimensionless constant.  Write $T=\eta_T R/c$ for a dimensionless constant $\eta_T$. Then,
\be
\label{Egw}
E_{GW} \sim\left(\frac{\eta_Q^2}{\eta_T^5} \right) M c^2  . 
\ee
The emitted radiation is sensitive to the transition time $T$. Note the similarity of these results to the simple barrier penetration ones in \eqref{EMest}.

The appearance of the total rest energy $Mc^2$  in \eqref{Egw} is not a surprise --- $M$ is the only dimensionfull parameter in the analysis. The dimensionless factor in front reflects theoretical uncertainty. For the LIGO gravitational wave event arising from the merger of two $30M_\odot$ black holes the combination is about $1/20$. Despite its crudity, \eqref{Egw} shows that detectable gravitational radiation from tunneling transitions between FB states is not ruled out by simple considerations of scale. 

These three examples show that the quantum dynamics of the fuzzball formation process and the structure of the FB solutions are important for observations. This is in sharp contrast with the coarse-grained description of the dynamics and the end state sometimes found in the literature in which the collapsed object is regarded as some kind of average over the FB ensemble. 

\section{Discussion}
\label{conclusion}

We have argued that the observational signatures of the formation of fuzzballs in gravitational collapse are likely to be more pronounced than has hitherto been thought. The basic reason is that a given collapse history corresponds to one particular branch of the wave function and not some kind of average over an ensemble of quasi-classical histories. Individual fuzzball solutions differ significantly from black holes in the neighbourhood of the collapsed object. This means that sufficiently fine-grained observations probing that region should enable one to differentiate between fuzzballs and black holes and, eventually, discriminate between different fuzzball solutions. 

Observations of special interest include the gravitational wave patterns from mergers and in particular the details of the ringdown phase. But also observations of the GW emission during inspirals of EMRI's at future gravitational wave observatories as well as EHT observations \cite{Doeleman:2009te} of test particles moving on geodesics in the neighbourhood of collapsed objects offer great potential. 

We have also identified a novel kind of GW burst associated with the quantum transitions that are involved in the formation process of fuzzballs in gravitational collapse. The expected amplitudes of these depend strongly on the transition times and the differences in the quadrupole and higher moments they generate, but are potentially observable. Both classes of observables provide a strong motivation for advancing the modeling of FBs and to derive the theoretical predictions for the probability distributions of some observables.

Observing the outcome of gravitational collapses at large distances is what is important for testing the predictions of the picture sketched. But it is not the only way the collapsed configuration could be observed. We can imagine sending an observer (or some recording device) into the collapsed configuration once it has settled down. Such an observer will see the features of the classical fuzzball geometry realized in that collapse and not some average fuzzball geometry. Since there is no horizon there is no obvious notion of complementarity, and no argument for the observer to see a firewall. There is also no reason that the information in what is recorded could not get out to an observer at infinity.  

The essential ingredients of our discussion occur in many other situations as we described in \cite{Hartle:2015bna}. Eternal inflation supplies an example in which the analogies are explicit. For certain matter field Lagrangians coupled to gravity the no-boundary wave function predicts significant probabilities for eternally inflationary histories which develop large fluctuations on very large scales, well outside our horizon, rendering the universe highly inhomogeneous on the largest scales. But we cannot observe these fluctuations ---- our observations are restricted to one Hubble volume. To predict probabilities for statistical features of fluctuation spectra on observable scales one can coarse-grain or ignore fluctuations on very large scales \cite{Hartle:2010dq}. The coarse-grained histories relevant for our local observations are nearly homogeneous even though he fine-grained histories are not.

\bigskip

\noindent \textbf{ Acknowledgements }
\bigskip

\noindent 

We thank Tom Lemmens and Bert Vercnocke for useful discussions. We thank the Center for the Fundamental Laws of Nature at Harvard and the Dept of Physics at Columbia University for their hospitality during respectively the initial and final stages of this work.
This work of JH was supported in part by the US NSF under grant PHY15-04541. The work of TH is supported in part by the National Science Foundation of Belgium (FWO) grant G092617N, by the C16/16/005 grant of the KULeuven and by the ERC grant no. ERC-2013-CoG 616732 HoloQosmos.

\bigskip
\bigskip

\bibliography{Literature_Fuzz}

\providecommand{\href}[2]{#2}\begingroup\raggedright\begin{thebibliography}{10}

\bibitem{Hartle2008}
J.~Hartle, S.~Hawking, and T.~Hertog, {\it {The Classical Universes of the
  No-Boundary Quantum State}},  {\em Phys.Rev.} {\bf D77} (2008) 123537,
  [\href{http://arxiv.org/abs/0803.1663}{{\tt arXiv:0803.1663}}].

\bibitem{Hartle2007}
J.~B. Hartle, S.~Hawking, and T.~Hertog, {\it {No-Boundary Measure of the
  Universe}},  {\em Phys.Rev.Lett.} {\bf 100} (2008) 201301,
  [\href{http://arxiv.org/abs/0711.4630}{{\tt arXiv:0711.4630}}].

\bibitem{Hartle:2015bna}
J.~Hartle and T.~Hertog, {\it {Quantum transitions between classical
  histories}},  {\em Phys. Rev.} {\bf D92} (2015), no.~6 063509,
  [\href{http://arxiv.org/abs/1502.06770}{{\tt arXiv:1502.06770}}].

\bibitem{Bramberger:2017cgf}
S.~F. Bramberger, T.~Hertog, J.-L. Lehners, and Y.~Vreys, {\it {Quantum
  Transitions Through Cosmological Singularities}},
  \href{http://arxiv.org/abs/1701.05399}{{\tt arXiv:1701.05399}}.

\bibitem{Mathur:2009zs}
S.~D. Mathur, {\it {How fast can a black hole release its information?}},  {\em
  Int. J. Mod. Phys.} {\bf D18} (2009) 2215--2219,
  [\href{http://arxiv.org/abs/0905.4483}{{\tt arXiv:0905.4483}}].

\bibitem{Mathur:2005zp}
S.~D. Mathur, {\it {The Fuzzball proposal for black holes: An Elementary
  review}},  {\em Fortsch. Phys.} {\bf 53} (2005) 793--827,
  [\href{http://arxiv.org/abs/hep-th/0502050}{{\tt hep-th/0502050}}].

\bibitem{Skenderis:2008qn}
K.~Skenderis and M.~Taylor, {\it {The fuzzball proposal for black holes}},
  {\em Phys. Rept.} {\bf 467} (2008) 117--171,
  [\href{http://arxiv.org/abs/0804.0552}{{\tt arXiv:0804.0552}}].

\bibitem{Balasubramanian:2008da}
V.~Balasubramanian, J.~de~Boer, S.~El-Showk, and I.~Messamah, {\it {Black Holes
  as Effective Geometries}},  {\em Class. Quant. Grav.} {\bf 25} (2008) 214004,
  [\href{http://arxiv.org/abs/0811.0263}{{\tt arXiv:0811.0263}}].

\bibitem{Bena:2013dka}
I.~Bena and N.~P. Warner, {\it {Resolving the Structure of Black Holes:
  Philosophizing with a Hammer}},  \href{http://arxiv.org/abs/1311.4538}{{\tt
  arXiv:1311.4538}}.

\bibitem{Mathur:2008kg}
S.~D. Mathur, {\it {Tunneling into fuzzball states}},  {\em Gen. Rel. Grav.}
  {\bf 42} (2010) 113--118, [\href{http://arxiv.org/abs/0805.3716}{{\tt
  arXiv:0805.3716}}].

\bibitem{Kraus:2015zda}
P.~Kraus and S.~D. Mathur, {\it {Nature abhors a horizon}},  {\em Int. J. Mod.
  Phys.} {\bf D24} (2015), no.~12 1543003,
  [\href{http://arxiv.org/abs/1505.05078}{{\tt arXiv:1505.05078}}].

\bibitem{Bena:2015dpt}
I.~Bena, D.~R. Mayerson, A.~Puhm, and B.~Vercnocke, {\it {Tunneling into
  Microstate Geometries: Quantum Effects Stop Gravitational Collapse}},  {\em
  JHEP} {\bf 07} (2016) 031, [\href{http://arxiv.org/abs/1512.05376}{{\tt
  arXiv:1512.05376}}].

\bibitem{Gibbons:2013tqa}
G.~W. Gibbons and N.~P. Warner, {\it {Global structure of five-dimensional
  fuzzballs}},  {\em Class. Quant. Grav.} {\bf 31} (2014) 025016,
  [\href{http://arxiv.org/abs/1305.0957}{{\tt arXiv:1305.0957}}].

\bibitem{Saxena:2005uk}
A.~Saxena, G.~Potvin, S.~Giusto, and A.~W. Peet, {\it {Smooth geometries with
  four charges in four dimensions}},  {\em JHEP} {\bf 04} (2006) 010,
  [\href{http://arxiv.org/abs/hep-th/0509214}{{\tt hep-th/0509214}}].

\bibitem{Balasubramanian:2006gi}
V.~Balasubramanian, E.~G. Gimon, and T.~S. Levi, {\it {Four Dimensional Black
  Hole Microstates: From D-branes to Spacetime Foam}},  {\em JHEP} {\bf 01}
  (2008) 056, [\href{http://arxiv.org/abs/hep-th/0606118}{{\tt
  hep-th/0606118}}].

\bibitem{Denef:2007yt}
F.~Denef, D.~Gaiotto, A.~Strominger, D.~Van~den Bleeken, and X.~Yin, {\it
  {Black Hole Deconstruction}},  {\em JHEP} {\bf 03} (2012) 071,
  [\href{http://arxiv.org/abs/hep-th/0703252}{{\tt hep-th/0703252}}].

\bibitem{Hartle:1992as}
J.~B. Hartle, {\it Space-time quantum mechanics and the quantum mechanics of
  space-time},  in {\em Gravitation and quantizations, Proc. Les Houches Summer
  School}, pp.~0285--480, 1992.
\newblock \href{http://arxiv.org/abs/gr-qc/9304006}{{\tt gr-qc/9304006}}.

\bibitem{Har93a}
J.~B. Hartle, {\em The Quantum Mechanics of Closed Systems}.
\newblock Directions in General Relativity Vol 1, Cambridge University Press,
  1993.

\bibitem{Giddings:2016btb}
S.~B. Giddings and D.~Psaltis, {\it {Event Horizon Telescope Observations as
  Probes for Quantum Structure of Astrophysical Black Holes}},
  \href{http://arxiv.org/abs/1606.07814}{{\tt arXiv:1606.07814}}.

\bibitem{Giddings:2016tla}
S.~B. Giddings, {\it {Gravitational wave tests of quantum modifications to
  black hole structure -- with post-GW150914 update}},  {\em Class. Quant.
  Grav.} {\bf 33} (2016), no.~23 235010,
  [\href{http://arxiv.org/abs/1602.03622}{{\tt arXiv:1602.03622}}].

\bibitem{Cardoso:2016oxy}
V.~Cardoso, S.~Hopper, C.~F.~B. Macedo, C.~Palenzuela, and P.~Pani, {\it
  {Gravitational-wave signatures of exotic compact objects and of quantum
  corrections at the horizon scale}},  {\em Phys. Rev.} {\bf D94} (2016), no.~8
  084031, [\href{http://arxiv.org/abs/1608.08637}{{\tt arXiv:1608.08637}}].

\bibitem{Abedi:2016hgu}
J.~Abedi, H.~Dykaar, and N.~Afshordi, {\it {Echoes from the Abyss: Evidence for
  Planck-scale structure at black hole horizons}},
  \href{http://arxiv.org/abs/1612.00266}{{\tt arXiv:1612.00266}}.

\bibitem{TheLIGOScientific:2016src}
{\bf Virgo, LIGO Scientific} Collaboration, B.~P. Abbott {\em et~al.}, {\it
  {Tests of general relativity with GW150914}},  {\em Phys. Rev. Lett.} {\bf
  116} (2016), no.~22 221101, [\href{http://arxiv.org/abs/1602.03841}{{\tt
  arXiv:1602.03841}}].

\bibitem{Audley:2017drz}
H.~Audley {\em et~al.}, {\it {Laser Interferometer Space Antenna}},
  \href{http://arxiv.org/abs/1702.00786}{{\tt arXiv:1702.00786}}.

\bibitem{Doeleman:2009te}
S.~Doeleman {\em et~al.}, {\it {Imaging an Event Horizon: submm-VLBI of a Super
  Massive Black Hole}},  \href{http://arxiv.org/abs/0906.3899}{{\tt
  arXiv:0906.3899}}.

\bibitem{Hartle:2016tpo}
J.~Hartle and T.~Hertog, {\it {One Bubble to Rule Them All}},
  \href{http://arxiv.org/abs/1604.03580}{{\tt arXiv:1604.03580}}.

\bibitem{Hartle:2010dq}
J.~Hartle, S.~W. Hawking, and T.~Hertog, {\it {Local Observation in Eternal
  inflation}},  {\em Phys. Rev. Lett.} {\bf 106} (2011) 141302,
  [\href{http://arxiv.org/abs/1009.2525}{{\tt arXiv:1009.2525}}].

\end{thebibliography}\endgroup
\bibliographystyle{JHEP}

\end{document}